\documentclass[aps,prx,twocolumn,superscriptaddress]{revtex4-2}
\usepackage{amsmath}
\usepackage{graphicx}

\begin{document}


\title{Spin flop and crystalline anisotropic magnetoresistance in CuMnAs}


\author{M. Wang}
\author{C. Andrews}
\affiliation{School of Physics and Astronomy, University of Nottingham, University Park, Nottingham NG7 2RD, United Kingdom}
\author{S. Reimers}
\affiliation{School of Physics and Astronomy, University of Nottingham, University Park, Nottingham NG7 2RD, United Kingdom}
\affiliation{Diamond Light Source, Harwell Science and Innovation Campus, Didcot OX11 0DE, United Kingdom}
\author{O. Amin}
\author{P. Wadley}
\author{R. P. Campion}
\author{S. F. Poole}
\author{J. Felton}
\author{K. W. Edmonds}
\email{Kevin.Edmonds@nottingham.ac.uk}
\author{B. L. Gallagher}
\author{A. W. Rushforth}
\author{O. Makarovsky}
\affiliation{School of Physics and Astronomy, University of Nottingham, University Park, Nottingham NG7 2RD, United Kingdom}
\author{K. Gas}
\author{M. Sawicki}
\affiliation{Institute of Physics, Polish Academy of Sciences, Aleja Lotnikow 32/46, PL-02668 Warsaw, Poland}
\author{D. Kriegner}
\affiliation{Institute of Physics, Czech Academy of Science, Cukrovarnick\'{a} 10, 162 00 Praha 6, Czech Republic}
\author{J. Zub\'{a}\v{c}}
\affiliation{Institute of Physics, Czech Academy of Science, Cukrovarnick\'{a} 10, 162 00 Praha 6, Czech Republic}
\affiliation{Faculty of Mathematics and Physics, Charles University, Ke Karlovu 3, 121 16 Prague 2, Czech Republic}
\author{K. Olejn\'{i}k}
\author{V. Nov\'{a}k}
\author{T. Jungwirth}
\affiliation{Institute of Physics, Czech Academy of Science, Cukrovarnick\'{a} 10, 162 00 Praha 6, Czech Republic}
\author{M. Shahrokhvand}
\author{U. Zeitler}
\affiliation{High Field Magnet Laboratory (HFML-EMFL), Radboud University, Toernooiveld 7, 6525ED Nijmegen, The Netherlands}
\author{S. S. Dhesi}
\author{F. Maccherozzi}
\affiliation{Diamond Light Source, Harwell Science and Innovation Campus, Didcot OX11 0DE, United Kingdom}



\date{\today}

\begin{abstract}
We report magnetic field induced rotation of the antiferromagnetic N\'{e}el vector in epitaxial CuMnAs thin films. Firstly, using soft x-ray magnetic linear dichroism spectroscopy as well as magnetometry, we demonstrate spin-flop switching and continuous spin reorientation in films with uniaxial and biaxial magnetic anisotropies, respectively, for applied magnetic fields of the order of 2~T. The remnant antiferromagnetic domain configurations are determined using x-ray photoemission electron microscopy. Next, we show that the N\'{e}el vector reorientations are manifested in the longitudinal and transverse anisotropic magnetoresistance. Dependences of the electrical resistance on the orientation of the N\'{e}el vector with respect to both the electrical current direction and the crystal symmetry are identified, including a weak 4th order term evident at high magnetic fields. The results provide characterization of key parameters including the anisotropic magnetoresistance coefficients, magnetocrystalline anisotropy and spin-flop field in epitaxial films of tetragonal CuMnAs, a candidate material for antiferromagnetic spintronics.
\end{abstract}


\maketitle

\section{Introduction}
In antiferromagnetic (AF) materials, the atomic magnetic moments alternate in direction to produce zero net magnetization. The consequent absence of magnetic stray fields and relative insensitivity to external fields offer advantages for certain memory applications \cite{jungw16,baltz18}, but also pose a challenge for the writing and reading of information. Recent works have shown that electric currents can be used to set the AF spin axis in crystalline materials, including CuMnAs and Mn$_2$Au, where the spin sublattices are space-inversion partners \cite{zelez14,wadley16,grzyb17,wadley18,bodnar18,meinert18,bodnar19,chen19}. While these observations open up opportunities to use AF materials in magnetic memory devices, a better understanding of the electrical readout mechanism is required. The rotation of the AF spin axis results in an anisotropic magnetoresistance (AMR) effect where the resistance depends on the relative orientation of the spin axis and current direction.  

A collinear AF material is characterized by a N\'{e}el vector $\mathbf{L} = (\mathbf{m_1}-\mathbf{m_2})$, where $\mathbf{m_1}$ and $\mathbf{m_2}$ are the magnetizations of each spin sublattice. An external magnetic field $H$ applied perpendicular to $\mathbf{L}$ competes against the strong internal exchange field $H_e$ which couples the sublattices. As a result, only a small canting of the magnetic moments into the field direction occurs when $H \ll H_e$, resulting in a small net magnetization $\mathbf{M} = (\mathbf{m_1}+\mathbf{m_2})$. A more significant reorientation can occur at the so-called spin-flop transition, from an AF state with $\mathbf{L}$ aligned parallel to the external field, to a state with $\mathbf{L}$ perpendicular, but with individual moments canted into the field direction. This is detected as a step-like increase in the magnetic susceptibility in bulk materials with uniaxial anisotropy, but is hard to detect in thin films due to the typically very small canting angle leading to a very weak magnetic signal.

AMR in AF materials has been explored only recently \cite{shick10,park11,song12,marti14,krieg16,krieg17,fina14,wang14,galce16,lu18,smejkal17}. In hexagonal MnTe films, a clear $\cos2\theta$ dependence on the angle $\theta$ between the current and the external field was demonstrated, due to the spin-flop rotation of the N\'{e}el vector \cite{krieg16}. Studies of the canted AF semiconductor Sr$_2$IrO$_4$ have shown a crystalline contribution to AMR, with four-fold symmetry, originating from changes in the equilibrium electronic structure induced by the rotation of the magnetic moments \cite{fina14,wang14,lu18}. In the case of the collinear antiferromagnet CuMnAs, understanding of the anisotropic magnetoresistance and magnetic anisotropy is crucially important for the design of memory devices based on the reorientation of the N\'{e}el vector and interpretation of their electrical readout signal. Moreover, due to the predicted dependence of the electronic band structure on N\'{e}el vector orientation in CuMnAs and related materials \cite{bodnar18,smejkal17}, a significant crystalline AMR effect may be expected, \emph{i.e.} a dependence of the electrical resistance on the directions of the current and the N\'{e}el vector with respect to crystalline symmetry \cite{krieg17}.

Here, we demonstrate magnetic field induced spin rotation and spin-flop behavior in epitaxial films of tetragonal CuMnAs, using x-ray magnetic linear dichroism (XMLD) spectroscopy as well as integral magnetometry. XMLD is the dependence of the x-ray absorption cross-section on the relative orientation of the x-ray linear polarization axis and the magnetization axis \cite{vdl86}. Unlike x-ray magnetic circular dichroism (XMCD), XMLD is an even function of the magnetic moment and can be detected equally in ferromagnets and antiferromagnets, and is widely used for, \emph{e.g.} AF domain imaging \cite{scholl00,nolting00},\cite{grzyb17,wadley18,bodnar18a}. We then show that the signature of the spin flop or spin reorientation can also be detected electrically. This allows us to determine the crystalline and non-crystalline contributions to the AMR in this material, of importance for interpreting and optimizing the electrical readout in AF memory devices.

\section{Sample growth and magnetic characterization}

\begin{figure}
\includegraphics[width=8.6cm]{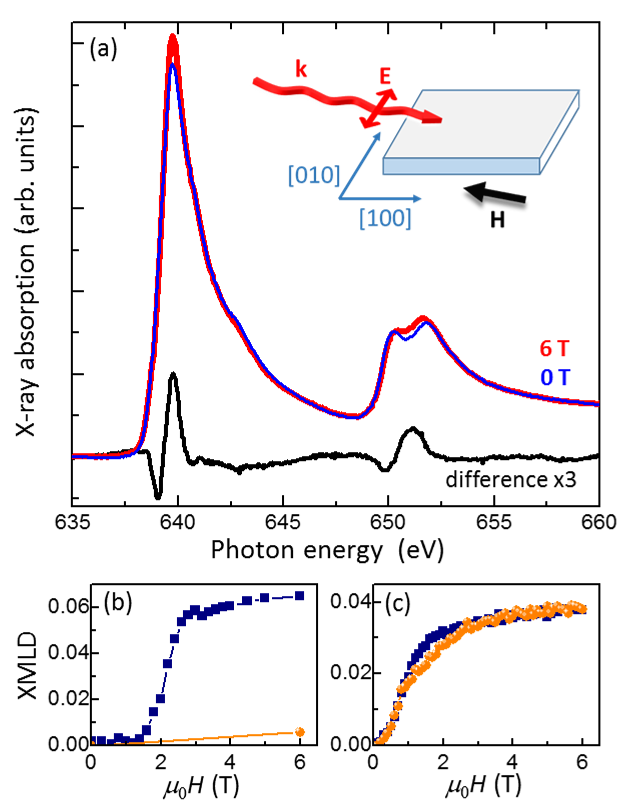}
\caption{\label{fig1}(a) Mn $L_{2,3}$ x-ray absorption spectra for 20~nm CuMnAs, in zero field (blue), 6~T field (red), and the difference (black), at 200~K. The x-ray incidence $\mathbf{k}$ and external magnetic field $\mathbf{H}$ are at $15^\circ$ to the sample surface, while the x-ray linear polarization $\mathbf{E}$ is along the CuMnAs [010] in-plane axis, as illustrated in the inset. The CuMnAs [100] axis is projected along the beam direction. (b,c) XMLD signal versus magnetic field for (b) 20~nm and (c) 50~nm CuMnAs films.  Blue squares and orange circles are for the in-plane projection of the field, $H_x$, along the [100] and [010] axes, respectively.}
\end{figure}

The tetragonal CuMnAs films, with thickness between 10~nm and 50~nm, were grown on GaP(001) substrates using molecular beam epitaxy. CuMnAs is lattice-matched to GaP through a $45^\circ$ rotation, such that the CuMnAs [100] axis is aligned with the GaP [110] \cite{wadley13}. For thin ($\le$ 20~nm) CuMnAs / GaP(001) layers, a uniaxial in-plane easy axis along the substrate [110] direction has been demonstrated \cite{wadley15,saidl17}, similar to well-known ferromagnet / III-V systems such as Fe/GaAs(001) \cite{bayre12}. Thicker films show a biaxial AF domain structure with easy axes along the CuMnAs [110] and [1\={1}0] axes \cite{wadley18}. 

Mn $L_{2,3}$ edge x-ray absorption spectroscopy measurements were performed on beamline I06-1 at Diamond Light Source, at a sample temperature of 200~K. The samples showed negligible XMCD even at the highest field applied of 6~T, indicating very small canting of the magnetic moments due to their strong AF coupling. The XMLD measurements were performed with the x-ray beam and the external magnetic field both at $15^\circ$ to the plane of the film [see Fig. 1(a), inset], with the x-ray linear polarization along the CuMnAs [010] in-plane direction (\emph{i.e.}, the substrate [1\={1}0] direction). The absorption spectra for a 20~nm film are shown in Fig. 1(a). A significant difference is observed between spectra obtained at low and high external magnetic fields, as shown by the difference spectrum in Fig. 1(a).

The difference spectrum in Fig. 1(a) is of comparable magnitude, relative to the absorption edge, to the XMLD spectrum previously observed in ferromagnetic Mn compounds \cite{freeman06,meinert11}. Also, comparable spectra are observed when the absorption is detected with both total electron yield (probing depth $\sim$ 3~nm) and fluorescence yield (probing depth $>$ 10~nm), ruling out a pure surface effect. Figure 1(b,c) shows the field-dependence of the XMLD signal, defined here as the peak-to-peak of the difference between spectra obtained in a magnetic field and in zero field.  Distinct behaviors are observed for 20~nm and 50~nm thick CuMnAs films. For 20~nm CuMnAs, the XMLD signal shows a sharp onset at around 1.5~T and a plateau at around 2.5~T for magnetic field projected along the [100] direction, while negligible XMLD is observed for field projected along the [010] direction. For the 50~nm film, similar behaviour is seen for both [100] and [010] directions, with an initial quadratic rise followed by saturation for fields above around 2~T.

The observed behavior is consistent with the N\'{e}el vector reorientation expected for a system with competing uniaxial and biaxial magnetic anisotropies. For the thinner film with dominant uniaxial anisotropy, the $\mathbf{L}$ vector undergoes a spin-flop transition into an axis perpendicular to the applied magnetic field. In this experimental geometry, $\mathbf{L}$ is perpendicular to the x-ray polarization at low fields, and parallel to it at high fields, resulting in a spectral change due to XMLD. For the thicker film, biaxial magnetic anisotropy is dominant, resulting in a so-called continuous spin-flop transition \cite{kliemt17} as the $\mathbf{L}$ vector rotates continuously into an axis perpendicular to the external field.

\begin{figure}
\includegraphics[width=8.6cm]{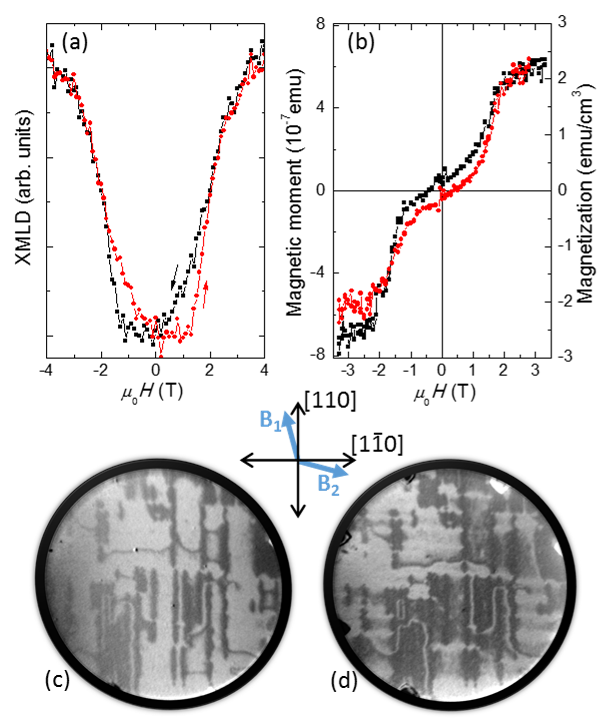}
\caption{\label{fig2}(a) XMLD hysteresis loop for a 10~nm CuMnAs film at 200~K. (b) SQUID hysteresis loop for a 10~nm CuMnAs film at 2~K, after removal of the substrate diamagnetic signal. (c,d) XPEEM images, with 30~$\mu$m field-of-view, of the antiferromagnetic domain structure in a 50~nm CuMnAs film at room temperature. The sample is at remanence after applying 7~T magnetic fields in the directions B1 and B2 shown in the inset. Dark / light domains correspond to a N\'{e}el vector parallel / perpendicular to the x-ray polarization,  which is along the CuMnAs [110] axis.}
\end{figure}

The XMLD hysteresis loop for a 10~nm CuMnAs film is shown in Fig. 2(a). At high magnetic fields, the XMLD signal is symmetric with respect to the field, as expected from its dependence on the square of the sublattice magnetic moment \cite{vdl86}. The hysteresis behaviour observed in the vicinity of the spin-flop transition indicates the formation of multidomain states in the thin CuMnAs film \cite{bogda07}.

To further confirm and quantify the spin-flop behavior, the field dependence of the magnetization $M(H)$ along the uniaxial easy axis of a 10~nm CuMnAs film was determined at 2~K using a Quantum Design MPMS superconducting quantum interference device (SQUID) magnetometry system \cite{sawicki11}. To \emph{in situ} compensate the much larger magnetic response of the GaP substrate, the sample was mounted between two abutting 8~cm long strips cut from another (equivalent) 2" GaP wafer, and careful calibration procedures were performed \cite{gas19}. The compensational sample holder assembled for these measurements is presented in Fig. 4(d) in ref. \cite{gas19}. The results [Fig. 2(b)] indicate a magnetization due to the spin canting of $(2.0\pm0.5)$~emu/cm$^3$ under 2~T field, where the uncertainty mostly arises from the subtraction of the substrate background signal. Using the local magnetic moment of $(3.6\pm0.2)$~$\mu_B$ per Mn atom in tetragonal CuMnAs obtained from neutron diffraction \cite{wadley13}, this corresponds to a canting angle of around 0.1-0.2$^\circ$. From this the exchange field can be estimated to be $\mu_0 H_e = (700\pm200)$~T. The spin-flop field $\mu_0 H_{sf}\approx$~2~T therefore corresponds to a magnetic anisotropy field of the order of $H_{sf}^2/H_e\approx$~5~mT. 

Figures 2(c) and (d) show images of the antiferromagnetic domain configuration in a 50~nm CuMnAs film obtained using x-ray photoemission electron microscopy, in remanence after applying 7~T magnetic fields at approximately $15^\circ$ from the CuMnAs [110] [Fig. 2(c)] and [1\={1}0] [Fig. 2(d)] biaxial easy axes. In both cases, the film is in a multidomain state after removing the external magnetic field, with a domain structure which is consistent with previous observations of similar CuMnAs films \cite{wadley18}. Similar features can be identified in each image, albeit with a significant remnant effect, \emph{i.e.} a preponderance of light domains in Fig. 2(c) and dark domains in Fig. 2(d). This indicates that the magnetization process in the biaxial CuMnAs films likely proceeds through N\'{e}el vector reorientation within the multiple domains as well as domain wall motion. 

\section{Anisotropic magnetoresistance}

\begin{figure}
\includegraphics[width=8.6cm]{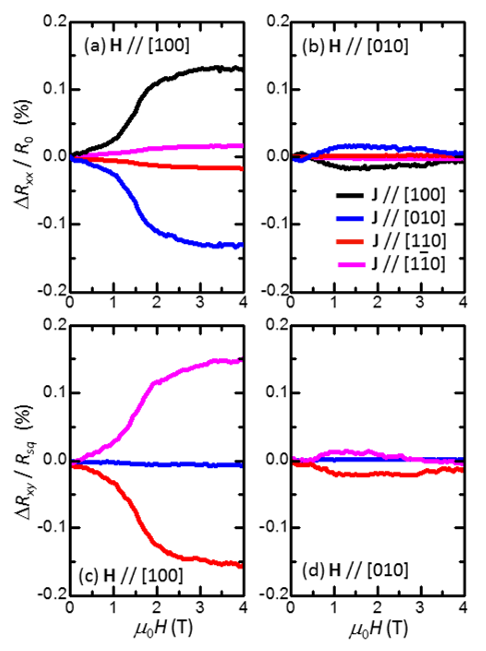}
\caption{\label{fig3}Longitudinal (a,b) and transverse (c,d) magnetoresistances for 10~nm thick CuMnAs at 4~K, with magnetic field applied along the [100] (a,c) and [010] (b,d) crystal axes, for various current directions.}
\end{figure}

DC magnetotransport measurements were performed on Hall bar devices fabricated from the CuMnAs films. Figure 3 shows data for the 10~nm thick CuMnAs film at 4~K, for devices with current channels along the [100], [010], [110] and [1\={1}0] crystalline directions. Magnetic fields were applied in the plane of the film at an angle $\theta$ to the current direction. Figure 3(a,b) show the normalized longitudinal component of the resistivity tensor [$(R_{xx} - R_{xx}(H=0))/R_{xx}(H=0)$, where $R_{xx}$ is the longitudinal resistance] as a function of magnetic field. A step-like change in the resistance is observed between 1~T and 2~T for fields applied along the easy axis [Fig. 3(a)], while a much weaker magnetoresistance is observed for field perpendicular to the easy axis [Fig. 3(b)].

Figure 3(c,d) shows the normalized transverse component of the resistivity tensor [$(R_{xy} - R_{xy}(H=0))/R_{sq}$, where $R_{sq}$ is the sheet resistance of $\approx 20~\Omega$ per square at zero field]. Again, a step-like change is observed for field applied along the easy axis. For the longitudinal and transverse resistances, the step-like behavior is observed for current parallel / perpendicular and at $45^\circ$ / $225^\circ$ to the magnetic field, respectively. This is consistent with the standard phenomenology of AMR \cite{mcguire75}. Including terms dependent on the crystalline anisotropy in a single crystal with twofold and fourfold symmetry components, the longitudinal and transverse components of AMR can be written \cite{rushf07}:

\begin{align}
AMR_{xx} &= (R_{xx} - R_{ave})/R_{ave} \nonumber \\
&= C_I \cos(2\phi) +C_U \cos(2\psi) \nonumber \\ 
&+ C_C \cos(4\psi) +C_{IC} \cos(4\psi - 2\phi)
\end{align}
\begin{align}
AMR_{xy} &= R_{xy}/R_{sq} \nonumber \\
&= C_I \sin(2\phi) - C_{IC} \sin(4\psi - 2\phi)
\end{align}

where $\phi$ and $\psi$ are the angles of the magnetization (for a ferromagnet) or the N\'{e}el vector (for an AF material) with respect to the current direction and the [100] crystalline axis, respectively. $R_{ave}$ is the longitudinal resistance averaged over a full rotation in the plane of the film. $C_I$, $C_U$, $C_C$ and $C_{IC}$ are phenomenological AMR coefficients corresponding to a non-crystalline term, twofold and fourfold crystalline terms, and a crossed crystalline / non-crystalline term, respectively.

\begin{figure}
\includegraphics[width=8.6cm]{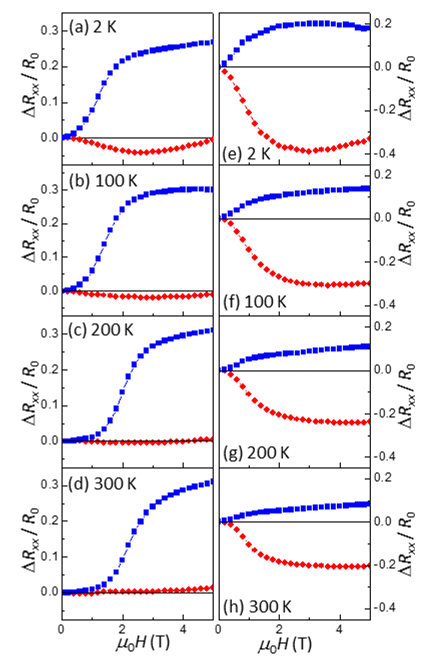}
\caption{\label{fig4}Longitudinal magnetoresistances at 2~K, 100~K, 200~K and 300~K for (a-d) 20~nm CuMnAs and (e-h) 50~nm CuMnAs films. The current is along the [100] and the magnetic field is applied along the [100] (blue squares) and [010] (red circles) directions.}
\end{figure}

The longitudinal magnetoresistance for 20~nm and 50~nm CuMnAs films at various temperatures are shown in Fig. 4, for current along the [100] and magnetic fields along the [100] and [010] directions. Uniaxial spin-flop behavior and continuous spin reorientation are observed for the 20~nm and the 50~nm films respectively, consistent with the XMLD field-dependence shown in Fig. 1(b,c). For the 20~nm film, with decreasing temperature, the spin-flop field decreases and a small negative magnetoresistance is observed for field along [010]. This indicates an increasing importance of the biaxial magnetic anisotropy with decreasing temperature, consistent with the expected strong dependence of biaxial anisotropy fields on sublattice magnetization \cite{callen66}. 

\begin{figure}
\includegraphics[width=8.6cm]{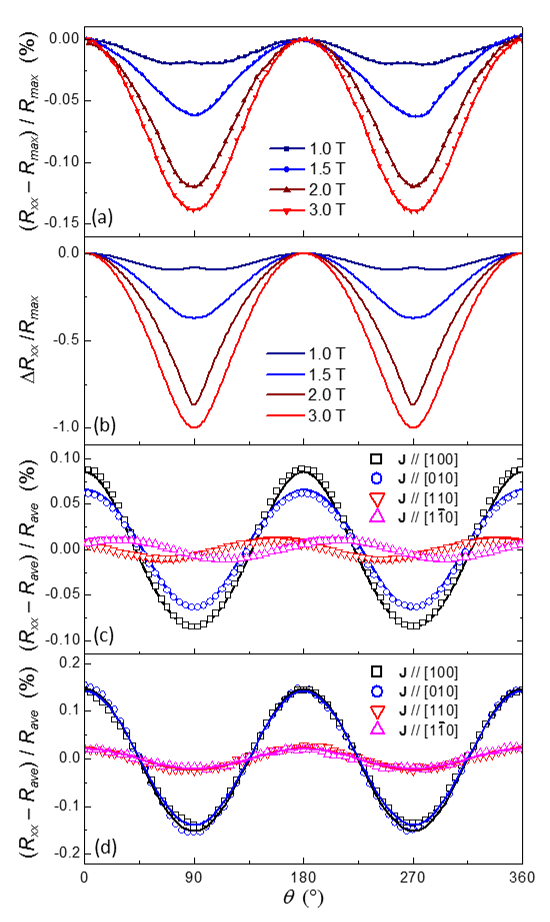}
\caption{\label{fig5}Resistance vs. angle $\theta$ between the magnetic field and current directions at 4~K. (a) 10~nm CuMnAs, current along the [010] direction, magnetic fields of magnitude 1.0~T, 1.5~T, 2.0~T and 3.0~T. (b) Calculated anisotropic magnetoresistance for a uniaxial antiferromagnet with a Gaussian distribution of spin-flop fields centered on 1.6~T. (c) 10~nm CuMnAs, current along the [100], [010], [110] and [1\={1}0] crystal directions, magnetic field of magnitude 5~T. (d) 50~nm CuMnAs, current along the [100], [010], [110] and [1\={1}0] crystal directions, magnetic field of magnitude 5~T. The lines in (c) and (d) are fits as described in the text.}
\end{figure}

The longitudinal resistance versus angle $\theta$ is shown in Fig. 5(a), for a 10~nm CuMnAs film at 4~K with magnetic fields comparable or smaller than $H_{sf}$. The current is along the [010] direction. With decreasing external field, the oscillations in the resistance become smaller and more anharmonic. This can be ascribed to the increasing importance of the magnetic anisotropy, resulting in the N\'{e}el vector only partially reorienting from its zero-field direction. The observed behavior can be simulated within a simple single-domain model including a uniaxial magnetic anisotropy. This simplified model agrees well with the observed dependence on both the magnitude and direction of the external magnetic field, including the appearance of additional features in the angle-dependence for $\mu_0 H=$1~T, as shown in Fig. 5(b).

At higher fields, the longitudinal resistance follows a $\cos2\theta$ dependence, as expected from equation (1) with $C_I, C_U, C_{IC} \gg C_C$ when the N\'{e}el vector aligns fully perpendicular to the field (\emph{i.e.}, $\phi = \theta$). However, as shown in Fig. 5(c,d) for 10~nm and 50~nm films, the observed $R_{xx}(\theta)$ depends strongly on the direction of the current, demonstrating the importance of the crystalline contributions to the AMR. For currents along the [110] and [1\={1}0] directions, the AMR is much reduced and for the 10~nm film is phase shifted by around $\pm25^\circ$. By fitting the field rotation data we obtain the AMR coefficients $C_I = (4.1\pm0.1)\times10^{-4}$, $C_U = (1.0\pm0.2)\times10^{-4}$ and $C_{IC} = (3.5\pm0.1)\times10^{-4}$ for the 10~nm film, and $C_I = (8.3\pm0.1)\times10^{-4}$, $C_U \approx 0$ and $C_{IC} = (6.2\pm0.1)\times10^{-4}$ for the 50~nm film. In both cases the $C_{IC}$ term is comparable in magnitude to the $C_I$ term. Therefore, these two terms largely cancel each other out for current along the [110] and [1\={1}0] directions, and the AMR is strongly suppressed. The non-negligible uniaxial term $C_U$ for the 10~nm film accounts for the difference in the AMR magnitude between the [100] and [010] current directions, and the phase shift for the [110] and [1\={1}0] directions. It is likely that this uniaxial crystalline AMR has the same origin as the uniaxial magnetic anisotropy giving rise to the spin flop, \emph{e.g.} anisotropic growth initiated at the III-V semiconductor surface.

\begin{figure}
\includegraphics[width=8.6cm]{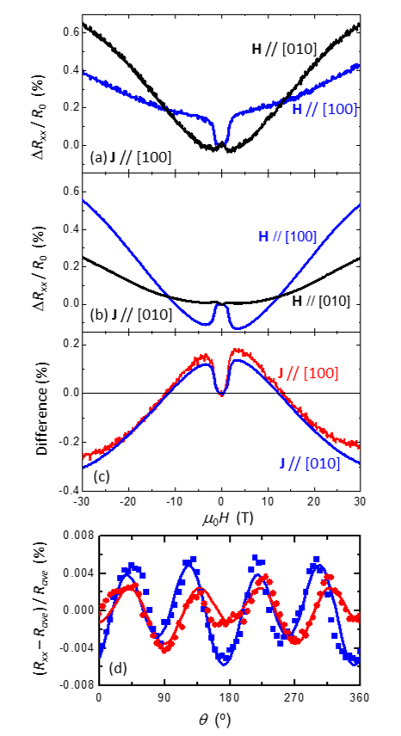}
\caption{\label{fig6}(a,b) Magnetoresistance for 10~nm CuMnAs under high magnetic fields at 4~K, for current along the [100] and [010] directions, and applied magnetic field perpendicular and parallel to the current. (c) Difference between the magnetoresistances for parallel and perpendicular orientations of current and magnetic field, for currents along the [100] and [010] directions. (d) Angle dependence of resistance for current along [010] at 13~T (squares) and for current along [100] at 14~T (circles). The lines are fits of the form $(a\cos2\theta+b\cos4\theta)$ with $a$ and $b$ as fit parameters.}
\end{figure}

The magnetoresistance at higher magnetic fields is shown in Fig. 6(a-c), for a 10~nm CuMnAs film at 4~K. In addition to the AMR, a positive magnetoresistance is observed for all directions of magnetic field and current. Moreover, the highest resistance occurs when the current and the magnetic field are perpendicular at high fields. This is opposite to the case at low fields where the highest resistance is when the field and current are coaxial. We ascribe the high field behaviour to a geometrical magnetoresistance effect, not directly related to the magnetic order, which is commonly observed in thin conducting films \cite{pippard}. Field rotation measurements at the crossover between low-field and high-field regimes, shown in Fig. 6(d), indicate a weak AMR with a dominant fourfold symmetry. For the rotation data at 5~T [Fig. 5(c)], the fourfold AMR term $C_C$ could not be distinguished due to the much larger two-fold AMR terms [see eq. (1)]. In the crossover regime [Fig. 6(d)], the twofold AMR terms are nearly cancelled by the geometrical magnetoresistance effect. The remaining fourfold AMR has resistance minima for magnetic fields along the $\langle110\rangle$ crystalline axes.

\section{Discussion and summary}
The magnetotransport effects shown in Figs. 3 to 6 provide a practical experimental tool for the determination of spin reorientation in antiferromagnetic domains and magnetic anisotropies in thin CuMnAs AF films. They also provide a direct measurement of the AMR coefficients and point to the importance of crystalline AMR terms in these epitaxial AF materials. The measured AMR corresponds to a fraction of a percent of the sheet resistance, comparable to the size of the electrical readout signals observed in the first demonstrations of current-induced switching in CuMnAs microdevices \cite{wadley16,grzyb17,wadley18}.

More recently, resistive signals of of the order of 10-100 percent have been observed in the same CuMnAs films as used in the present study. These signals are switchable with unipolar current pulses or polarization-independent optical pulses, even in magnetic fields much larger than $H_{sf}$ \cite{kaspar}. The present study provides confirmation that the large switching signals are not linked to a net reorientation of the N\'{e}el vector and the corresponding AMR. A recent magnetostatic imaging study has instead linked the large unipolar switching signal to a current-induced fragmentation to nanoscale AF domain textures, and subsequent relaxation towards an equilibrium domain configuration \cite{wornle}. In principle, however, so far unidentified structural effects may also contribute or even govern these large effects. Future systematic experiments in magnetic field might, therefore, not only contribute to our understanding of the current-induced N\'{e}el vector switching and AMR but also of these high-resistive switching signals in CuMnAs.

\begin{acknowledgments}
This work was supported by the Engineering and Physical Sciences Research Council Grant No. EP/P019749/1, the EU FET Open RIA Grant No. 766566, the Ministry of Education of the Czech Republic Grants LM2015087 and LNSM-LNSpin, and the Grant Agency of the Czech Republic Grant No. 19-28375X. We thank the Diamond Light Source for the allocation of beam time under Proposals SI11846 and SI20793. The high magnetic field study was supported by HFML-RU/NWO-I, member of the European Magnetic Field Laboratory (EMFL) and by EPSRC via its membership to the EMFL (grant no. EP/N01085X/1). We thank Prof. Amalia Patane for providing access to a high field cryostat at the University of Nottingham. PW acknowledges support from the Royal Society through a University Research Fellowship.
\end{acknowledgments}

\bibliography{spinflop}

\end{document}